**Is Quantum Mechanics Incompatible with Newton's First Law?**

**Mario Rabinowitz**

**Abstract** Quantum mechanics (QM) clearly violates Newton's First Law of Motion (NFLM) in the quantum domain for one of the simplest problems, yielding an effect in a force-free region much like the Aharonov-Bohm effect. In addition, there is an incompatibility between the predictions of QM in the classical limit, and that of classical mechanics (CM) with respect to NFLM. A general argument is made that such a disparity may be found commonly for a wide variety of quantum predictions in the classical limit. Alternatives to the Schrödinger equation are considered that might avoid this problem. The meaning of the classical limit is examined. Critical views regarding QM by Schrödinger, Bohm, Bell, Clauser, and others are presented to provide a more complete perspective.

**Keywords** Aharonov-Bohm effects · Quantum mechanical classical limit · Non-locality · Free particle in a box · Newton's first law in quantum mechanics · Wave packets · Expectation values

**1 Introduction**

Quantum Mechanics (QM) is considered to be a theory that applies to every particular case in both the microscopic and macroscopic realms of our physical world. Yet QM has fared badly in the quantum gravity realm [18, 19], and there is no extant theory after eight decades of effort [22, 24]. In the case of the macroscopic classical realm, it is generally expected that quantum results should correspond to classical results in the limit of large quantum number $n$, or equivalently in the limit of Planck's constant $h \to 0$.

One object of this paper is to show that in the force-free region inside a box (infinite square well potential), there is an effect on a particle much in common with [1, 2], and similar effects, except for the absence of $h$ and no need to invoke the canonical momentum.

________

M. Rabinowitz
Armor Research, 715 Lakemead Way, Redwood City, CA 94062-3922, USA
e-mail: Mario715@gmail.com
Another object of this paper is to show that the classical limit of a quantum mechanical calculation for such a force-free particle differs from the classical result. Since there is disagreement in this simplest possible case, this raises a basic question regarding the universality of quantum mechanics.

Newtonian laws of motion are the basis of classical mechanics (CM) as expounded in Isaac Newton's Principia [16]. The first two laws are:

**Newton's First Law of Motion** Every body continues in its state of rest or of uniform motion in a right line [straight ahead], unless it is compelled to change that state by forces impressed upon it.

**Newton's Second Law of Motion** The change of motion [of a body] is proportional to the motive force impressed; and is made in the direction of the right [straight] line in which that force is impressed.

Newton's first law is a special case of his second law since the absence of a force leaves the body in its original state of uniform motion. We shall see that Newton's second law can be derived from Schrödinger's equation, which was set up to yield Newton's laws of motion in the classical limit. So it is doubly troubling that in the classical limit, quantum mechanics appears incompatible with Newton's first law.

It is generally held and sanctioned that wave packets are excellent representations of classical macroscopic bodies. Bohm [4] says:

> Such solutions [plane waves] actually represent an abstraction never realized in practice, because all real waves are bounded in one way or another. A wave packet more closely represents what happens in a real experiment. . . . We conclude therefore that in all phenomena not requiring a description in terms of distances as small as a de Broglie length, the use of de Broglie wave packets leads to exactly the same results as does classical mechanics.

In addition to showing the incompatibility of QM with NFLM in the quantum realm, this paper will examine the generally accepted and commonly taught dictum that wave packets composed of small wavelength (high energy and hence large quantum number) plane waves give the same result as classical mechanics. This is usually the limit in which the finite size

-2-

and the internal structure of the wave packet can be ignored. We will mainly be considering wave packets for a force-free particle in zero potential; however the results appear to be more generally applicable. The correspondence principle of QM requires that the expectation (average) values of wave packet variables equal the classical particle variables, or QM is not a complete theory of nature.

## 2 The Meaning of the Classical Limit

A traditional classical limit of QM is attained by taking the limit $h \to 0$. This may seem peculiar, since $h$ is not a dimensionless number, but has units of energy multiplied by time such as Joule-sec, erg-sec, etc. So $h$ can be small in one set of units or large in another. Let's see what this means implicitly.

From the point of view of QM, matter behaves classically, when a particle of momentum $p$ has a small spatial change in wavelength $\lambda = (h/p)$ during interaction or course of observation

$$1 >> \left|\frac{\partial \lambda}{\partial x}\right| = \left|\frac{\partial}{\partial x}\left(\frac{h}{p}\right)\right| = \left|\left(\frac{-h}{p^2}\right)\frac{\partial p}{\partial x}\right| \Rightarrow h << \frac{p^2}{\left(\frac{\partial p}{\partial x}\right)} \qquad (2.1)$$

Therefore Eq. (2.1) implies that

$$\left[\frac{h}{\left(\frac{p^2}{dp/dx}\right)}\right] \to 0. \qquad (2.2)$$

This is properly dimensionless, and what is meant by $h \to 0$.

The argument that the classical limit corresponds to high energies and hence large quantum number $n$, can be ambiguous. This is because $h$ and $n$ often enter in as a product in the QM energy and the limit $h \to 0$ and $n \to \infty$ can be undetermined:

$$hn \xrightarrow[h \to 0 \,\&\, n \to \infty]{} ?$$

However, the limit of $n \to \infty$ can be approached without consideration of energy. This alternate perspective is that the classical limit occurs when the spreading of a wave packet can be neglected in the context of a given experiment or observation. If a Gaussian wave packet for a mass m is prepared at time $t = 0$ with root mean square width $A$, a fundamental result of QM is that its width $x$ will automatically increase with time without any external influence as given by:

$$\Delta x = A\left[1+\left\{\frac{(h/2\mathbf{p})t}{2mA^2}\right\}^2\right]^{1/2}. \tag{2.3}$$

For a free particle of momentum $mv$ confined to a region of width 2a, the de Broglie relation implies that

$$m = \frac{h}{\mathbf{l}v} = \frac{h}{(4a/n)v} = \frac{hn}{4av}. \tag{2.4}$$

Substituting Eq. (2.4) into Eq. (2.3):

$$\Delta x = A\left[1+\left\{\frac{(h/2\mathbf{p})t}{2mA^2}\right\}^2\right]^{1/2} = A\left[1+\left\{\frac{avt}{\mathbf{p}A^2 n}\right\}^2\right]^{1/2} \xrightarrow[n\to\infty]{} \Delta x = A. \tag{2.5}$$

So a wave packet prepared in a very high quantum number state has a width that hardly spreads because it is essentially time independent (and $h$ independent). So, this is consistent with $h$? 0 and $n$? 8, or $h$? 0 and $m$? 8. The Dirac equation has an equally troubling mass dependent motion for a free relativistic electron in free space: zitterbewegung. The center of a wave packet is analogous to the center of mass of a macroscopic body though the two may differ in the classical limit. The center of a wave packet corresponds to that position where waves of different quantum number (hence different energy $E$) remain approximately in phase. Mathematically this can be stated as $?S/?E = 0$, so the center is where the phase $S$ is an extremum as a function of energy of the wave packet that is expanding in time.

Bohr [8] introduced the correspondence principle for QM to yield the classical limit. This correspondence is broadly accommodated if any one of three cases can do the job:

1. In the limit of large quantum number $n$.

2. The system is described by a large set of quantum numbers or combinations of eigenfunctions.

3. Or both cases 2 and 3.

Bohr thus delimits QM to those solutions that indeed correspond to CM in the classical limit, but then this is a correction in hindsight. Even then, the classical limit of Schrodinger's wave mechanics is at best statistical, i.e., Schrodinger's wave mechanics does not reduce to deterministic classical mechanics. Of course, it is not clear what the classical limit is for difficult cases like the three or more-body problem, where the solution is not tractable and the motion appears chaotic. Some processes thought to be purely and uniquely in the quantum realm like tunneling, can with proper modeling also exist in the classical realm as shown by Cohn and Rabinowitz [11]. Because it is so fundamental, and the classical implications of Newton's first law of motion are easily tractable, it seems ideally suited for testing the validity of QM in describing the classical realm.

**3 Newton's Second Law from Schrödinger's Equation**

It is well known that Newton's second law can be derived from QM as presented in many textbooks [20]. It is one example of Ehrenfest's theorem that quantum expectation values correspond to classical observables. The starting point in the derivation is the definition for the expectation value of the momentum $p$ in terms of the momentum operator $\left(\frac{-i\hbar}{2\pi}\nabla\right)$, where $\tau$ is the volume:

$$\langle p \rangle = \int_{-\infty}^{\infty} \psi^* \left(\frac{-i\hbar}{2\pi}\nabla\right) \psi \, d\tau$$

$$\frac{d}{dt}\langle p \rangle = \frac{d}{dt}\int_{-\infty}^{\infty} \psi^* \left(\frac{-i\hbar}{2\pi}\nabla\right) \psi \, d\tau = -i\frac{\hbar}{2\pi}\left[\int_{-\infty}^{\infty} \psi^* \nabla \frac{\partial \psi}{\partial t} d\tau + \int_{-\infty}^{\infty} \frac{\partial \psi^*}{\partial t} \nabla \psi \, d\tau\right]$$

$$= -\int_{-\infty}^{\infty} \psi^* \nabla\left[\frac{(\hbar/2\pi)^2}{2m}\nabla^2 \psi + V\psi\right] d\tau + \int_{-\infty}^{\infty} \left[\frac{(\hbar/2\pi)^2}{2m}\nabla^2 \psi^* + V\psi^*\right]\nabla \psi \, d\tau \quad (3.1)$$

$$= -\int_{-\infty}^{\infty} \psi^*[\nabla(V\psi) - V\nabla\psi]d\tau = -\int_{-\infty}^{\infty} \psi^*[\nabla V]\psi \, d\tau \equiv \langle -\nabla V \rangle = \langle F \rangle$$

Equation (3.1) is Newton's second law as obtained by substituting the time independent term of Schrödinger's equation for the time dependent term, and integrating twice by parts.

Although it appears to be a fundamental derivation, it is really just an illustration of the internal consistency of QM. This result should not be surprising as minus the gradient of the potential energy $(-\nabla V)$ is the force ($F$) operator. But the final step of (3.1) could just as well have been the starting point for the correspondence of these QM expectation values to the classical values. Neither going from the first step of the derivation to the last step, nor just starting from the last step can be considered to be a derivation from first principles.

The statement that CM is derivable from QM doesn't prove that the classical limit of quantum mechanics is more fundamental than classical mechanics. It only demonstrates the self-consistency of QM, and the epistemological compatibility of the two theories. Thus despite the above derivation, it will be surprising to see that QM may be incompatible with Newton's first law, since the first law is just a special case of the second when $F = 0$ (implying that the momentum $p$ is constant). This is furthermore unexpected, because the free particle wave packet group velocity, $v_g = p/m$ equals the classical particle's velocity $v$. In the next section we will examine the consequences of a free particle wave packet in the classical limit.

## 4 Quantum Analysis of the Free Particle

4.1 Pure Quantum State and Classical Expectation Values

Non-relativistic quantum mechanics is embodied in the Schrödinger wave equation:

$$\frac{-(h/2\pi)^2}{2m}\nabla^2\psi + V\psi = i(h/2\pi)\frac{\partial}{\partial t}\psi, \tag{4.1}$$

where $\psi$ is the wave function of a particle of mass $m$, with potential energy $V$. In the case of constant $V$, we can set $V = 0$ as only differences in $V$ are physically significant. A solution of (4.1) for the one-dimensional motion of a free particle of $n$th state kinetic energy $E_n$ is:

$$\psi = b_n e^{i2\pi x/\lambda} e^{-i2\pi E_n t/h} = b_n e^{i2\pi\left(\frac{x}{\lambda} - \nu_n t\right)}, \tag{4.2}$$

where the wave function $\psi$ travels along the positive $x$ axis with wavelength $\lambda$ and frequency $\nu = E/h$ and phase velocity $v = \nu\lambda$.

We shall be interested in the time independent solutions. The following forms are equivalent:

$$y_n = b_n e^{i2px/l} = b_n \cos(2px/l) + i\sin(2px/l), \quad n = 1, 2, 3, \ldots \quad (4.3)$$
$$= b_n \sin(npx/2a - np/2)$$

where we can consider the particle to be in an infinite square well potential with perfectly reflecting walls at $x = -a$, and $x = +a$, so that $(1/2)\prime = 2a$. (*a* can be arbitrarily large, but needs to be finite so that the normalization coefficient is non-zero.) To illustrate the presence of symmetric and anti-symmetric waves, the solutions (4.3) can also be written as

$$y_{n\,symmetric} = y_{ns} = b_n \cos\left[\frac{cpx}{2a}\right] \text{ for any odd integer } c = 1, 3, 5, \ldots, \quad (4.4a)$$

where the nth symmetric eigenfunction is given by $n_s = c$.

The symmetric and antisymmetric eigenfunctions alternate such that

$$y_{n\,antisymmetric} = y_{nas} = b_n \sin\left[\frac{spx}{2a}\right] \text{ for any even integer } s = 2, 4, 6, \ldots, \quad (4.4b)$$

where the nth antisymmetric state is given by $n_{as} = s$.

We normalize the wave functions to yield a total probability of $1 = \int_{-a}^{a} y^* y \, dx = \int_{-a}^{a} |y|^2 \, dx$ of finding the particle in the region -a to +a, and find

$$y_{ns} = \frac{1}{\sqrt{a}} \cos\left[\frac{cpx}{2a}\right], \text{ and} \quad (4.5)$$

$$y_{nas} = \frac{1}{\sqrt{a}} \sin\left[\frac{spx}{2a}\right], \quad (4.6)$$

where the normalization is independent of n and of the nature of the symmetry.

Let us find the expectation values <x> and <x²> for the free particle in the nth state. In general

$$\langle x^k \rangle = \int_{-a}^{a} y^* x^k y \, dx = \int_{-a}^{a} x^k |y|^2 \, dx, \text{ for } k = 1, 2, 3, \ldots. \quad (4.7)$$

Since $|y|^2$ is symmetric **here** for both $y_{ns}$ and $y_{nas}$, $x^k |y|^2$ is antisymmetric in the interval -a to +a, because $x^k$ is antisymmetric for odd k. Thus without having to do the integration we know that $\langle x^k \rangle = 0$ for all odd k, and in particular $\langle x \rangle = 0$ for the nth state. Doing the integration for the expectation value of <x> and <x²> for the nth state:

$$\langle x \rangle_{QM} = \int_{-a}^{a} y^* x y \, dx = \int_{-a}^{a} x |y|^2 \, dx = 0 \quad (4.8)$$

$$\left\langle x^{2}\right\rangle_{QM} =\int_{-a}^{a} \psi^{*} x^{2}\psi dx=\int_{-a}^{a} x^{2}|\psi|^{2} dx = a^{2}\left[\frac{1}{3}-\frac{1}{p^{2}n^{2}}\right]. \tag{4.9}$$

Let us compare these values with the corresponding classical values. The classical probability P is uniform for finding a classical free particle in the region -a to +a. Normalizing the classical probability, $1=\int_{-a}^{a} Pdx = P(2a) \Rightarrow P = \frac{1}{2a}$. As for the quantum case, classically $<x^k> = 0$ for all odd k because P is an even function. The classical expectation value of $<x>$ and $<x^2>$ are

$$\left\langle x\right\rangle_{ClassicalMechanics}=\left\langle x\right\rangle_{CM} = \int_{-a}^{a} Pdx = \int_{-a}^{a}\frac{x}{2a}dx=0 \tag{4.10}$$

$$\left\langle x^{2}\right\rangle_{ClassicalMechanics}=\left\langle x^{2}\right\rangle_{CM} = \int_{-a}^{a} Pdx = \int_{-a}^{a}\frac{x^{2}}{2a}dx = \frac{a^{2}}{3}. \tag{4.11}$$

As one might expect, $\left\langle x^{2}\right\rangle_{QM} \xrightarrow{n\to\infty} \left\langle x^{2}\right\rangle_{CM} = \frac{a^{2}}{3}$ for a pure state n.

## 4.2 The Meaning of $\left\langle x^{2}\right\rangle_{QM}$ for Low Quantum Numbers

The result $x$ QM = 0 of (4.8) is the same as $x$ CM = 0 of (4.10), and both are in accord with Newton's First Law of Motion (NFLM). As expected, this means that in moving with a constant velocity between the walls of a box, a particle spends an equal amount of time on either side of the box and hence the expectation value for finding it, is at the center of the box. However, we saw from (4.9) $\left\langle x^{2}\right\rangle_{QM} = a^{2}\left[\frac{1}{3}-\frac{1}{p^{2}n^{2}}\right]$ for a particle in a perfectly reflecting box of length 2a between walls. At low quantum number n, this is smaller than the classical value $\left\langle x^{2}\right\rangle_{CM} = \frac{a^{2}}{3}$ of (4.11). This tells that not only does the particle spend an equal time on either side of the origin, but that the particle spends more time near the center of the box independent of the length *a*. [This would not be the case for an unconfined particle.] Since we can make the length *a* extremely large, this effect is due to quantum mechanical non-locality of the presence of the walls making itself felt near the center of the box because it does not go away with large *a*.

This is a violaton of NFLM because the particle must slow down in the region of the origin even though there is a force on it only at the walls. The particle cannot both be going at a constant velocity between the walls, slow down near the center, and speed up again as it goes toward the opposite wall even if the walls are extremely far apart. Therefore in this example, we have a quantum action on a particle even where there is no force. This is a simpler case than the Aharonov-Bohm, Aharonov-Casher, and similar effects, has many of the same elements, and may be even more intrinsic to QM. It is noteworthy that unlike such effects, it is independent of Planck's constant *h*.

4.3 Expectation Values for Mixed Quantum States

The QM solution for a pure state is not a good representation for a classical particle which is better represented as a mixture of states i.e. by a wave packet. So let's first observe the generality of adding wave functions. Given that

$$\frac{-(h/2\pi)^2}{2m}\nabla^2\psi_n + V\psi_n = i(h/2\pi)\frac{\partial}{\partial t}\psi_n. \tag{4.12}$$

$$\Rightarrow \sum_{n=1}^{\infty}\left[\frac{-(h/2\pi)^2}{2m}\nabla^2\psi_n + V\psi_n\right] = \sum_{n=1}^{\infty} i(h/2\pi)\frac{\partial}{\partial t}\psi_n. \tag{4.13}$$

$$\Rightarrow \frac{-(h/2\pi)^2}{2m}\nabla^2\sum_{n=1}^{\infty}\psi_n + V\sum_{n=1}^{\infty}\psi_n = i(h/2\pi)\frac{\partial}{\partial t}\sum_{n=1}^{\infty}\psi_n. \tag{4.14}$$

Therefore if each $\psi_n$ is a solution of the Schrödinger Eq. (4.1), then any sum including an infinite sum of $\psi_n$ is also a solution of the Schrödinger equation. This conclusion is so general that there is no restriction that each $\psi_n$ represent a pure state. So $\psi_n = \sum_{j=1}^{n}\psi_j$ is equally valid as next shown.

$$\sum_{n=1}^{\infty}\sum_{j=1}^{n}\left[\frac{-(h/2\pi)^2}{2m}\nabla^2\psi_j + V\psi_j - i(h/2\pi)\frac{\partial}{\partial t}\psi_j\right] = 0, \tag{4.15}$$

since each term in the sum is equal to 0.

$$\Rightarrow \sum_{n=1}^{\infty}\sum_{j=1}^{n}\left[\frac{-(h/2\pi)^2}{2m}\nabla^2\psi_j + V\psi_j\right] = \sum_{n=1}^{\infty}\sum_{j=1}^{n} i(h/2\pi)\frac{\partial}{\partial t}\psi_j \tag{4.16}$$

$$\Rightarrow \frac{-(h/2\boldsymbol{p})^2}{2m}\nabla^2\sum_{n=1}^{\infty}\sum_{j=1}^{n}\boldsymbol{y}_j + V\sum_{n=1}^{\infty}\sum_{j=1}^{n}\boldsymbol{y}_j = i(h/2\boldsymbol{p})\frac{\P}{\P t}\sum_{n=1}^{\infty}\sum_{j=1}^{n}\boldsymbol{y}_j. \tag{4.17}$$

$$\Rightarrow \frac{-(h/2\boldsymbol{p})^2}{2m}\nabla^2\sum_{n=1}^{\infty}\boldsymbol{y}_n + V\sum_{n=1}^{\infty}\boldsymbol{y}_n = i(h/2\boldsymbol{p})\frac{\P}{\P t}\sum_{n=1}^{\infty}\boldsymbol{y}_n. \tag{4.18}$$

Q.E.D. since Eq. (4.18) yields Eq. (4.15).

We now make some general observations regarding any Even Function *E(x)*, and any Odd Function *O(x)*. The product *E(x)* times *O(x)* is an odd function. The product *E(x)* times *E(x)* is an even function. The product *O(x)* times *O(x)* is an even function. Therefore the product of *O(x)* times *E(x)* times *E(x)* is an odd function; and the product of *O(x)* times *O(x)* times *E(x)* is an even function. The integral over symmetric limits of an odd function is = 0. The integral over symmetric limits of an even function may be = 0, >0, or <0.

Let $\boldsymbol{y}_{Ei}(x)$ be an Even Wave Function, and $\boldsymbol{y}_{Oi}(x)$ be an Odd Wave Function. Then

$$\boldsymbol{y}_{Ei}(x)\boldsymbol{y}_{Oi}(x) = \boldsymbol{y}_{Oj}(x) \Rightarrow \sum_i \boldsymbol{y}_{Ei}(x)\boldsymbol{y}_{Oi}(x) = \sum_j \boldsymbol{y}_{Oj}(x), \tag{4.19}$$

where $\boldsymbol{y}_{Oj}(x)$ is a different function than the wave function $\boldsymbol{y}_{Oi}(x)$.

Similarly

$$\sum_j \boldsymbol{y}_{Oi}(x)\boldsymbol{y}_{Oi}(x) = \sum \boldsymbol{y}_{Ej}(x), \text{ and } \sum \boldsymbol{y}_{Ei}(x)\boldsymbol{y}_{Ei}(x) = \sum_j \boldsymbol{y}_{Ej}(x), \tag{4.20}$$

where $\boldsymbol{y}_{Ej}(x)$ is a different function than the wave function $\boldsymbol{y}_{Ei}(x)$.

For a wave packet that consists entirely of a sum of even functions or entirely of a sum of odd functions that satisfy Schrödinger's equation the odd-moment (odd k) position expectation value is 0:

$$\langle x^k \rangle = \int_{-a}^{a} x^k \left[\sum_i |\boldsymbol{y}_{Ei}(x)\boldsymbol{y}_{Ei}(x)|\right]^2 dx = 0, \text{ and } \langle x^k \rangle = \int_{-a}^{a} x^k \left[\sum_i |\boldsymbol{y}_{Oi}(x)\boldsymbol{y}_{Oi}(x)|\right]^2 dx = 0, \tag{4.21}$$

since $x^k$ is odd.

Now consider the cross terms for which $x^k \boldsymbol{y}_{Ei}(x)\boldsymbol{y}_{Oi}(x) = E(x)$, since O(x)E(x)O(x) =

$O(x)O(x) = E(x)$.  In taking $y^*y$ there will be even terms times even terms and odd terms times odd terms, all of which are even.  But most importantly there will be odd terms times even terms all of which are odd; and which when multiplied by (odd k) $x^k$ are even.

In the next section, a detailed argument will be presented that piecewise integration of the cross products yields:

$$\int_{-a}^{a} \sum_i \left( x^k \cos\left[\frac{u\mathbf{p}x}{2a}\right] \sin\left[\frac{s\mathbf{p}x}{2a}\right] \right)_i dx = \sum_i \int_{-a}^{a} \left( x^k \cos\left[\frac{u\mathbf{p}x}{2a}\right] \sin\left[\frac{s\mathbf{p}x}{2a}\right] \right)_i dx \neq 0 \tag{4.22}$$

Therefore as we shall see in detail:

$$\langle x^k \rangle = \int_{-a}^{a} x^k \left[ \sum_i |\mathbf{y}_{Ei}(x)\mathbf{y}_{Oi}(x)| \right]^2 dx \neq 0, \tag{4.23}$$

where k is any odd integer, and in particular for k = 1, $\langle x \rangle_{QM} \neq 0$.

4.4  General Wave Packet To Represent A Classical Free Particle

Let us consider the most general wave function that can be used to represent a comprehensive wave packet for a free particle:

$$\Psi = B \sum_{n=1}^{N} b_n \sin\left(\frac{n\mathbf{p}x}{2a} - \frac{n\mathbf{p}}{2}\right) e^{-i2\mathbf{p}nt}, \tag{4.24}$$

where the sum is over all possible eigenfunctions each of quantum number n, where the highest quantum number N is a very large number that may approach infinity if necessary.  The normalization condition

$$1 = \int_{-\infty}^{\infty} \Psi^* \Psi \, dx \Rightarrow 1 > B^2 > 0, \tag{4.25}$$

insures that the total probability is equal to 1, of finding the free particle moving in the x direction with $-a < x < a$.  For finite $a$, the contribution to the integral is 0 for $|x| > a$.  To reduce the complexity of the analysis, let us only consider the quantum expectation value of the particle's position:

$$\langle x \rangle_{QM} = \int_{-a}^{a} \Psi^* x \Psi \, dx = B^2 \int_{-a}^{a} x \left[ \sum_{n=1}^{N} b_n \sin\left(\frac{n\mathbf{p}x}{2a} - \frac{n\mathbf{p}}{2}\right) \right]^2 dx. \tag{4.26}$$

Whatever conclusion we reach for $\langle x \rangle_{QM}$ will also be precisely true for $\langle x^k \rangle_{QM}$, where k is any odd integer, as $x^k$ is an odd function. The analysis for $\langle x^k \rangle_{QM}$, where k is any even integer is not included here because of its greater complexity. For even k such as $\langle x^2 \rangle_{QM}$, in the following analysis one must deal with the much larger set of even functions in the cross products and because $\langle x^2 \rangle_{QM}$ involves the quantum number n. Preliminary results indicate that for a wave packet $\langle x^2 \rangle_{QM} \neq \langle x^2 \rangle_{CM}$.

For the sake of clarity let us explicitly introduce the even cosine function (C) and the odd sine function (S) in the squared factor of Eq. (4.26):

$$\left[\sum_{n=1}^{N} \sin(\frac{n p x}{2a} - \frac{n p}{2})\right]^2 = \left[\sum_{c=1}^{N} \cos\left(\frac{c p x}{2a}\right) + \sum_{s=2}^{N+1} \sin\left(\frac{s p x}{2a}\right)\right]^2$$

$$= \sum \cos^2\left[\frac{c p x}{2a}\right] + \sum \sin^2\left[\frac{s p x}{2a}\right] + 2\left[\sum \cos\left(\frac{c p x}{2a}\right)\cos\left(\frac{(c+2) p x}{2a}\right) + \sum \sin\left(\frac{s p x}{2a}\right)\sin\left(\frac{(s+2) p x}{2a}\right) + \sum \cos\left(\frac{c p x}{2a}\right)\sin\left(\frac{s p x}{2a}\right)\right]$$

(4.27)

where the index $c$ runs over the odd integers, $c = 1, 3, 5, \ldots$, and the index $s$ runs over the even integers, $s = 2, 4, 6, \ldots$. The symbols $c$ and $s$ were chosen as mnemonics for the indices of the cosine and sine functions respectively. The constant coefficients $bn$ were chosen to be unity for clarity and their presence would not change the tenor of the argument. They can be viewed as a convenience for representing a wave packet with a limited number of eigenfunctions. The analysis here includes the possibility of an unlimited number of eigenfunctions as $N \to \infty$.

Note from Eq. (4.27) that only the last summation $\sum \cos\left(\frac{c p x}{2a}\right) \sin\left(\frac{s p x}{2a}\right)$ contributes to $\langle x \rangle_{QM}$ as each term of the other sums is even which times x yields an integration = 0. In shorthand representation the integral of Eq. (4.26) is left with:

$$\langle x \rangle_{QM (N+1)} = B^2 \int_{-a}^{a} x\left[C_1 \sum_{s=2}^{N+1} S_s + C_3 \sum_{s=2}^{N+1} S_s + C_5 \sum_{s=2}^{N+1} S_s + \ldots + C_N \sum_{s=2}^{N+1} S_s\right] dx,$$

(4.28)

where this summation ends with the (N+1)th eigenfunction $\sin\left(\frac{(N+1)\boldsymbol{p}x}{2a}\right)$, i.e. an even number of (N + 1) eigenfunctions are used to construct the wave packet. [As stated at the beginning of the previous paragraph C and S stand for cosine and sine in the last summation $\sum \cos\left(\frac{c\boldsymbol{p}x}{2a}\right)\sin\left(\frac{s\boldsymbol{p}x}{2a}\right)$, with corresponding indices.] Similarly if the summation has an odd number of eigenfunctions and only goes up to the Nth eigenfunction, $\cos\left(\frac{N\boldsymbol{p}x}{2a}\right)$:

$$\langle x \rangle_{QM\,(N)} = B^2 \int_{-a}^{a} x\left[C_1\sum_{s=2}^{N-1}S_s + C_3\sum_{s=2}^{N-1}S_s + C_5\sum_{s=2}^{N-1}S_s + \ldots + C_N\sum_{s=2}^{N-1}S_s\right]dx. \tag{4.29}$$

$$\langle x \rangle_{QM\,(N+1)} - \langle x \rangle_{QM\,(N)} = B^2 \int_{-a}^{a} x[C_1 S_{N+1} + C_3 S_{N+1} \ldots + C_N S_{N+1}]dx > 0 \tag{4.30}$$

Eq. (5.7) is dominated by the last term where the close match in frequencies produces a reinforcement resonance-like condition in amplitude:

$$\left|B^2\int_{-a}^{a}x[C_N S_{N+1}]dx\right| > \left|B^2\int_{-a}^{a}x[C_1 S_{N+1} + C_3 S_{N+1}\ldots + C_N S_{N-1}]dx\right| \Rightarrow \langle x \rangle_{QM\,(N+1)} > \langle x \rangle_{QM\,(N)}, \tag{4.31}$$

and $\langle x \rangle_{QM\,(N+1)} > 0$ since $B^2\int_{-a}^{a}x[C_N S_{N+1}]dx > 0$. But the last term alternates in sign and

$B^2\int_{-a}^{a}x[C_N S_{N-1}]dx < 0$. So $\langle x \rangle_{QM\,(N-1)} < 0$. Therefore $\langle x \rangle_{QM}$ alternates in sign, being positive when the last eigenfunction making up the wave packet is antisymmetric, and being negative when the last eigenfunction is symmetric. This implies that $x$ QM is indeterminate.

Normally a modest superposition of about 20 states is sufficient to produce a compact wave packet. This analysis indicates that even with an arbitrarily large number of states, a wave packet cannot yield the proper classical limit for the position expectation value of a free particle. This oscillation resembles zitterbewgung.

Since the classical expectation value <x>$_{CM}$ = 0 as given by (4.10), it has been shown here that for a wave packet the classical limit of quantum mechanics $|\langle x \rangle_{QM}| \neq 0$ cannot agree with the classical prediction and is also indeterminate when a wave packet contains odd and even functions. This is a general argument that should hold for both free particles and particles acted on by a symmetrical potential such as that of the simple harmonic oscillator. Since the Dirac and Klein Gordon equations reduce to the Schrodinger equation, one may expect them to have the same difficulty in the classical limit.

**5 Discussion**

To my knowledge, this is the first time Newton's first Law has been examined in this way with respect to the quantum mechanical prediction of a free particle's position expectation value. When others such as [15] examine Newton's First Law of Motion (NFLM) in the context of QM, they usually are considering the dispersion of the wave packet for a free particle which interestingly is mass dependent. Note that h is not present in $<x^k>$ where k is an odd integer such as $<x^1>$, nor in $<x^{k+1}>$ such as $<x^2>$. So one might expect no quantum effect here, except with respect to reflection. Perfect reflection is an idealization that implies instantaneous and dissipationless reflection. As such it does not occur in the quantum or the classical world. The limitations of real reflection can be minimized by making the length *a* sufficiently large, but this will not change the non-local quantum violation of NFLM in the force free region near the center of the box.

Since the reflection process was not modeled in the quantum calculation, it might be surprising in prospect that QM violates NFLM in the quantum realm. In retrospect, we can say that this is the way it should be because the wave function solutions give an increased probability of finding the particle in the region near the center. What is surprising is that the disparity between QM and CM in the quantum realm is fundamental to the Schrödinger equation, and does not begin to involve questions related to quantum electrodynamic (QED) differences between the quantum vacuum and the classical vacuum. For an unbounded free particle in QM, the plane waves have a continuum of eigenfunctions because there is no boundary constraint that limits them to integral quantum numbers. However there are no truly plane waves in nature because there are always boundaries. Similarly there are no monochromatic (monoenergetic) beam of particles. Yet the plane wave approximation is routinely and successfully made. The approximation of a particle in a box is at least as good. Since the Dirac and the Klein-Gordon equations reduce to the Schrödinger equation in the non-relativistic limit, one may expect the conclusions of this paper apply to them also. Similarly for Schrödinger's wave mechanics and Heisenberg's matrix mechanics, since they are equivalent. When the two approaches were shown to be equivalent, the outstanding mathematician David Hilbert reputedly said that physics is obviously far too difficult to be left to physicists, and mathematicians still think they are God's gift to science.

5.1 Alternative Wave Equations

It would not be surprising if an experimental confirmation is found for the force free deviation from NFLM in the quantum realm. The Aharonov-Bohm, and other force-free effects have been experimentally confirmed.

The linear superposition of states in quantum mechanics is one of the peculiarities of quantum mechanics that makes it distinct from classical mechanics. The disparity of the classical limit of $x$ QM for a wave packet with $x$ CM may lie in the absence of a proper formulation for making a wave packet. Or it may point to the need to modify the Schrödinger equation so that its predictions will be compatible with CM in the classical limit. There has been much work done related to nonlinear generalizations of quantum mechanics and devising experimental tests of theories governed by such equations [26]. The nonlinear Schrödinger equation frequently appears in such generalizations. To date there has been no strong experimental evidence to support the tested variations. Another possibility is that experiment in the classical realm will support the predictions of QM in the classical limit rather than the classical predictions, as is the case for superconductivity.

Bohm [5–7] has argued that classical physics does not emerge from quantum physics in the same way that classical mechanics emerges as an approximation of special relativity at small velocities; rather, classical physics exists independently of quantum theory and cannot be derived from it. The analysis in this paper is in accord with this view. However, the objections raised here apply equally to De Broglie-Bohm Pilot Wave QM [14] as they do to ordinary QM discussed in this paper. The addition of a Quantum Potential to the Schrödinger equation allows for a different interpretation of a particle guided by a quantum wave. However this addition does not change the solutions. Thus the classical limit disparities apply to both theories. Hopefully this will lead to a better theory. Furthermore both QM and the Bohm version of QM violate the Weak Equivalence Principle (WEP) that the motion of a body in a gravitational field is independent of its mass and hence also of its composition [18, 19]. (The WEP is the gravitational counterpart of NFLM, which they also violate.) Albert Einstein said that every theory is killed sooner or later . . . . But if the theory has good in it, that good is embodied and continued in the next theory.

## 5.2 Critical Views of Quantum Mechanics

It is not well known that J.S. Bell of Bell's theorem fame relating to quantum entanglement was quite concerned about QM. Bell [3] favored the Pilot wave model despite experimental success in showing that QM was in accord with his theorem. Bell [3] is quite critical of QM, as is Erwin Schrödinger. Bell quotes Schrödinger as saying, "If we have to go on with these damned quantum jumps, then I'm sorry I ever got involved." On reflecting on the progress of a quarter of a century of QM, Schrödinger [21] observes that QM cannot account for the particularity (clarity) of the world of experience from the macro to the micro, as compared with the indefiniteness of QM. Peres [17] a leading practitioner of QM evidenced his disdain for the Uncertainty Principle in his book on QM.

Even a foremost defender of orthodox QM, R.B. Griffiths [12, 13] says: "Wave function collapse can be assigned to the trash can of outmoded ideas, replaced by a consistent set of conditional probabilities." Griffiths echoes what many said before him. This is no small condemnation, as wave function collapse was invented by John von Neumann [25] to patch up inconsistencies between QM and the physical world that occur in the quantum realm.

The fact that Clauser [10] (who was instrumental in conducting the first experiments to test Bell's theorem in favor of QM) has also expressed grave doubts about QM, casts a double shadow on QM. Clauser starts out his Chapter:

> This article is dedicated to the memory of John Bell, whose work exerted a profound influence on my own life and professional career as an experimental physicist.
> In this article I attempt to recount as accurately as possible the important events in the development of one of the most profound results in physics of the twentieth century – Bell's Theorem.

Clauser goes on to critique not only QM but also quantum electrodynamics (QED) with respect to the groundbreaking work of Stroud and Jaynes [23]:

> Jaynes and his students found that (without QED) they could predict absorption of radiation, spontaneous and stimulated emission of radiation, the Lamb shift, and the black body radiation spectrum.
> Jaynes' work, however had crossed a magic line in the sand and obviously had to be viewed as heresy. Heretofore, it was a firmly held belief that all of these latter effects do require a quantization of the radiation field.

Credit must also be given to Timothy Boyer who created what is now called Stochastic Electrodynamics (SED). Boyer's new theory of random electrodynamics is a classical electron theory involving Newton's equations for particle motion due to the Lorentz force, and Maxwell's equations for the electromagnetic fields with point particles as sources. Boyer [9] introduced a background of random, classical fluctuating zero-point fields whose origin was in the initial stochastic processes of the big bang and are regenerated to the present. Boyer's theory of random electrodynamics is a classical theory that provides a link between classical theory with $h = 0$ and quantum electrodynamics.

## 6 Conclusion

This paper shows the intrinsic non-locality of quantum mechanics (QM) even in the simplest of cases such as a bounded free particle. This results in an effect much like the class of Aharonov-Bohm, Aharonov-Casher, and similar effects resulting in a violation of Newton's First Law of Motion (NFLM) in the quantum realm (low quantum numbers), and in the classical limit for wave packets. In the quantum realm, symmetry causes the expectation value of $x$ to be centered at $x = 0$, meaning that the particle spends equal time on both sides of the origin as in the classical case. However, the QM non-locality manifests itself in $\langle x^2 \rangle_{QM} < \langle x^2 \rangle_{CM}$ showing that the particle spends most of its time in the region near the origin. Thus the particle slows down as it approaches the origin, and speeds up as it moves towards the wall. The non-locality manifests itself by means of a high quantum number wave packet even in the classical limit.

Thus this paper presents a new challenge that quantum mechanics may not be a complete theory that applies to every particular case in both the microscopic and macroscopic realms of our physical world. A class of QM predictions appears to be incompatible with classical mechanics.

As shown in Sect. 3 the observation that Newton's first law [when there is no applied force the momentum is constant] is a special case of Newton's second law which is derivable from QM carries no special significance. It was designed to do so. Since macroscopic objects in the classical domain must be represented by quantum wave packets, we need to compare the predictions of QM for wave packets rather than pure state predictions in the classical limit with the predictions of CM. As shown in this paper, an appreciable difference between the two theories becomes manifest. Ultimately, the question of which is correct must be decided empirically.